\documentclass[12pt,preprint]{aastex}
\shorttitle{Mid-Infrared Spectroscopy of Disks around CTTS}
\shortauthors{Forrest et al.}

\begin{document}

\title{Mid-Infrared Spectroscopy of Disks around Classical T Tauri Stars}

\author{
W.J. Forrest\altaffilmark{1},
B. Sargent\altaffilmark{1},
E. Furlan\altaffilmark{2},
P. D'Alessio\altaffilmark{3},
N. Calvet\altaffilmark{4},
L. Hartmann\altaffilmark{4},
K.I. Uchida\altaffilmark{2},
J.D. Green\altaffilmark{1},
D.M. Watson\altaffilmark{1},
C.H. Chen\altaffilmark{5},
F. Kemper\altaffilmark{6},
L.D. Keller\altaffilmark{7},
G.C. Sloan\altaffilmark{2},
T.L. Herter\altaffilmark{2},
B.R. Brandl\altaffilmark{8},
J.R. Houck\altaffilmark{2},
D.J. Barry\altaffilmark{2},
P. Hall\altaffilmark{2},
P.W. Morris\altaffilmark{9},
J. Najita\altaffilmark{10},
and P.C. Myers\altaffilmark{4}}
\email{forrest@pas.rochester.edu}

\altaffiltext{1}
{Department of Physics and Astronomy, University of
Rochester, Rochester, NY 14627-0171}
\altaffiltext{2}
{Center for Radiophysics and Space Research,
Cornell University, Space Sciences Building, Ithaca,
NY 14853-6801}
\altaffiltext{3}
{Centro de Radioastronomia y Astrofisica, UNAM, Apartado
Postal 3-72 (Xangari), 58089 Morelia, Michoac\'an, Mexico}
\altaffiltext{4}
{Harvard-Smithsonian Center for Astrophysics,
60 Garden Street, Cambridge, MA 02138}
\altaffiltext{5}
{National Research Council Resident Research Associate,
Jet Propulsion Laboratory, M/S 169-506, California
Institute of Technology, 4800 Oak Grove Drive, Pasadena,
CA 91109}
\altaffiltext{6}
{Spitzer Fellow, Department of Physics and Astronomy, University
of California, Los Angeles, CA 90095-1562}
\altaffiltext{7}
{Department of Physics, Ithaca College, Ithaca, NY 14850}
\altaffiltext{8}
{Sterrewacht Leiden, P.O. Box 9513, 2300 RA Leiden,
Netherlands}
\altaffiltext{9}
{Spitzer Science Center/Infrared Processing and Analysis
Center, Calfornia Institute of Technology, Pasadena, CA 91125}
\altaffiltext{10}
{National Optical Astronomy Observatory, 950 North Cherry
Avenue, Tucson, AZ 85719}
\clearpage

\begin{abstract}
We present the first Spitzer Infrared Spectrograph\footnote[11]{The IRS
was a collaborative venture between Cornell University and Ball
Aerospace Corporation funded by NASA through the Jet Propulsion
Laboratory and the Ames Research Center.} observations of the
disks around classical T Tauri stars: spectra in the 5.2-30 $\mu$m
range of six stars.  The spectra are dominated by emission features from
amorphous silicate dust, and a continuous component from 5 to 8
$\mu$m that in most cases comprises an excess above the photosphere
throughout our spectral range. There is considerable variation in the
silicate feature/continuum ratio, which implies variations of
inclination, disk flaring, and stellar mass accretion rate. In most of
our stars, structure in the silicate feature suggests the presence of
a crystalline component. In one, CoKu Tau/4, no excess above the
photosphere appears at wavelengths shortward of the silicate features,
similar to 10 Myr old TW Hya, Hen 3-600, and HR 4796A. This indicates
the optically thick inner disk is largely absent.  The silicate
emission features with peaks at 9.7 and 18 $\mu$m indicate small dust
grains are present.  The extremely low 10-20 $\mu$m color temperature
of the dust excess, 135 K, indicates these grains are located more than 10
AU from the star.  These features are suggestive of gravitational
influence by planets or close stellar companions and grain growth in
the region within 10 AU of the star, somewhat surprising for a star this
young (1 Myr).
\end{abstract}


\section{Introduction}
Although \citet{lynden-bell74} suggested that T Tauri stars
were surrounded by accretion disks, it was not until the advent of
IRAS that observations over a wide wavelength range demonstrated that
the infrared excesses were due to dust disk emission \citep{rucinski85}.
Optically-thick models for disk emission powered mostly by accretion
\citep{kenyon95} or heated mostly by irradiation by the
central star \citep{adams87} were developed to explain the
broad features of T Tauri spectral energy distributions (SEDs).
\citet{kenyon87} pointed out that the overall SEDs could best
be explained if the dust was suspended to several scale heights in the
gas, resulting in a disk photosphere that is curved away from the
midplane, or ``flared''.  Flaring enhances the irradiation heating by
the central star, increasing mid- to far- infrared fluxes.  The outside-in
heating of disks by stellar radiation leads to warmer temperatures in
outer layers than near the midplane, which in turn predicts silicate
features in emission in optically thick disks \citep{calvet92}.
Because the disk heating affects its vertical structure, which in turn
changes the amount of stellar irradiation, self-consistent models have
been developed to predict SEDs in greater detail \citep{dalessio98,
dalessio99, dalessio01, chiang97}.

The IRS instrument on the Spitzer Space Telescope covers a crucial
wavelength range (typically corresponding to disk radii of one to a
few AU) for constraining disk structure, and provides an unparalleled
opportunity to address particle composition near the disk surface.  We
present our preliminary findings on a small sample of T Tauri stars in
the Taurus-Auriga dark clouds.  The 5\,--\,30 $\mu$m spectra are broadly consistent
with model expectations.  There is typically a strong 10 $\mu$m
silicate emission feature overlying a continuum well in excess of the
stellar photosphere.  There is evidence for grain processing in the
changing shape of this feature.  One of the objects, CoKu Tau/4,
shows no detectable excess shortward of 8 $\mu$m, but a rapidly rising
flux to longer wavelengths with 10 and 18 $\mu$m
silicate features in emission.  The implied dust temperatures are quite
low, leading us to interpret this as a disk in transition:  the region
within 10 AU has been largely cleared of observable, small, dust grains
while the outer disk remains.

\section{Observations}

We observed 20 classical T Tauri stars (SED Class II) in the Taurus dark
clouds during the 4 February 2004 Infrared Spectrograph
\citep[IRS:][]{houck04} campaign on the Spitzer Space Telescope
\citep{werner04}. From these we have selected six representative
stars (see Table 1).  Five of the stars were selected to not have a known
close binary companion.  GG Tau is in a more complex system.  The object
we observed, GG Tau Aa, has a close (0\,{\farcs}25) binary companion,
GG Tau Ab, which is also a classical T Tauri star with active accretion
\citep{hartigan03}.  Aa is a factor of 2 brighter than Ab in the K and L
bands \citep[2.2 and 3.6 $\mu$m;][]{white01}.  Our spectrum represents
Aa+Ab. \\
We present observations obtained with the Short-Low (SL; 5.2\,--\,14
$\mu$m, ${\lambda}/{\Delta}{\lambda}$ $\sim$ 90) and Long-Low 
(LL; 14\,--\,38 $\mu$m, ${\lambda}/{\Delta}{\lambda}$ $\sim$ 90) 
modules of the IRS.
FM Tau, CY Tau, and CoKu Tau/4 were observed in the IRS Staring mode
(two nod observations on the slit per target), while FN Tau, GG Tau, and IP Tau 
were observed using the IRS Spectral Mapping Mode (2$\times$3 steps of the
slit across each target). The details of how these modes were used are the same 
as described by \citet{uchida04}.  
% In the case of FN Tau, one of the nod positions in SL1 (the first order of the SL 
% module, 7.5\,--\,14 $\mu$m) was lower in flux by  $\sim$ 10 \% than the 
% other. The bonus order (7.4\,--\,8.7 $\mu$m), which was observed at the 
% same time as SL2 (the second order of the SL module, 5.2\,--\,8.7 $\mu$m),
% agreed in flux with the higher flux nod position so only that data is presented here.  
% The lower flux nod position agreed quite well in the shape of the spectrum.  
For the SL data shown here, the mispointing parallel to the slit was typically of the
order of 0.3 pixels; since our spectral reduction method requires that our science 
targets be placed at precisely the same position on the slit as the calibration star, 
such small mispointings do not affect the calibration of our spectra.

We reduced our spectra with the IRS team's Spectral Modelling,
Analysis, and Reduction Tool \citep[SMART;][]{higdon04} as described
by \citet{uchida04}. $\alpha$ Lac (A1 V) was used to calibrate our SL
and LL spectra; the resulting spectra are shown in Figures 1 and 2.
The only LL spectra presented here are the ones of FM Tau and
CoKu Tau/4.  We have compared the fluxes derived in this manner
for SL to published fluxes in the literature, primarily N band and
IRAS 12 $\mu$m fluxes, for about six non-variable
objects.  The agreement was always better than 10\%.  We have also
compared the IRAC 5.8 and 8 $\mu$m observations of TW Hya
to the SL spectra in \citet{uchida04}.  The agreement was
better than 10\%.  We conclude that over the flux range 30 mJy to 1.5
Jy the photometric accuracy of our SL calibration is at least as good
as 10\%. For any undetected, modest mispointing, our reported fluxes 
would be too low, but the shape of the true spectrum would be preserved.
Using the same calibrator star ($\alpha$ Lac), the LL spectrum is
typically above the SL spectrum at 14 $\mu$m, by factors of 1.2\,--\,1.4.
This mismatch is believed to be primarily due to an error in the early
pipeline which has been identified and is being corrected.
We believe the shape of the LL spectrum is basically correct.
The LL spectra in Figure 2 have been multiplied by 0.7 to match SL.

In addition to photometric accuracy, the accuracy of the shape of the
spectra is crucial.  We have addressed this issue in two ways.  Firstly, the
SL spectra of several objects reduced using the same hardware, software,
and procedures showed no evidence of excess emission.  Those spectra
were used to assess our photometric accuracy discussed above.  The shape 
of these SL spectra was in agreement, within the spectral errors, to that of 
the stellar photosphere, which was modelled as a blackbody with the effective 
temperature of the star (essentially a Rayleigh-Jeans tail here).  Secondly, 
the spectral error in each data point was assessed by comparing the spectra 
from the two independent nod positions for each object included in this paper.  
Half the absolute value of the difference is the standard deviation of the 
mean if the sample distribution is gaussian.  We find that these error bars 
agree with the scatter in the data points shown here.  Since we extract two 
data points per spectral resolution element, neighboring data points are not
statistically independent.  The scatter in neighboring data points
indicates the spectral noise.  Only features which significantly
exceed this scatter can be considered real.

\section{Discussion}

The spectra of FM Tau, GG Tau, and IP Tau (Fig. 1) are all similar.
From 5\,--\,8 $\mu$m, there is a smooth continuum that is shallower than
the Rayleigh-Jeans tail of the stellar photosphere.  In order to
estimate the stellar contribution to these spectra, a simple model for
the stellar photosphere was constructed.  The effective temperatures
were assumed to be those consistent with the spectral class (see Table
1).  While the extinction to the stars (Table 1) is small, the contribution
of the accretion disk to the flux at wavelengths shortward of 5 $\mu$m
is not precisely known.  Therefore, we have assumed the star provides
all of the flux in the K band 2MASS observation (see note for Table
1).  This will give an upper limit to the stellar flux in the 5\,--\,30
$\mu$m range.  The implied star sizes are typically 2 solar radii for
an assumed distance of 140 pc.  This is consistent with stars 1\,--\,3 Myr
old, typical of Taurus members.

For five of the six stars, the 5\,--\,8 $\mu$m continuum flux is well
above the stellar continuum.  This is a well known characteristic of the
accretion disks around young solar mass stars \citep{kenyon87, calvet92}.
The disk is optically thick and heated by direct and scattered stellar and
accretion-shock radiation, re-radiation from the extended disk atmosphere,
and conversion of gravitational potential energy to heat.  The temperature
in the disk increases with decreasing radius; at some point the temperature
is high enough ($\sim$ 1500 K) to evaporate the refractory dust.
The observed flux represents the sum over many annuli, with many
different temperatures.

Beyond 8 $\mu$m, there is a prominent emission feature which we
identify with emission from small silicate dust grains.  This is also a
well-known feature of these accretion disks \citep{calvet92, cohen85}.
The optically thin emission comes from dust suspended in the atmosphere
of the flared, optically thick accretion disk.  The flaring permits exposure
to more nearly direct starlight, promoting heating.  Since these small grains
absorb starlight more efficiently than they emit in the infrared, the
temperatures are elevated above that of a blackbody grain at the same
radius \citep{calvet92, chiang97}. The grain temperature is well above
the temperature in the optically thick accretion disk below, permitting the
appearance of emission features in the spectrum.  Such ``superheating'' has
been seen in the small silicate dust grains given off by comets in our solar
system \citep{gehrz92}.

The smoothest emission feature, which is also among the narrowest of the 
sample, is seen in FM Tau.  It peaks near 9.7 $\mu$m, and the shape is 
similar to that seen in ISM dust as exemplified by the heated dust in the 
Trapezium \citep{forrest75, gillett75}. There is a slight ``knee'' at 11.3 
$\mu$m which is not evident in the Trapezium spectrum.  The LL spectrum 
(Fig. 2) reveals a second emission feature, peaking near 18 $\mu$m.  We 
identify this with the bending-mode resonance in small amorphous silicate 
grains.  Its shape is very similar to that seen in the Trapezium 
\citep{forrest76a, forrest76b}. We conclude that the 8\,--\,30 $\mu$m
emission from the optically thin, superheated, flared disk comes primarily from
small silicate grains, similar to those found in the ISM.

The 9.7 $\mu$m silicate feature of IP Tau is very similar to FM Tau's.
The only noticeable difference is a more definite ``knee'' at 11.3 $\mu$m,
perhaps indicating a small amount of crystalline silicates.

The 9.7 $\mu$m silicate feature of GG Tau is also similar to FM Tau's.
It shows a more distinct ``knee'' near 11.3 $\mu$m than IP Tau.  It also has
relatively enhanced flux in the 8\,--\,9.7 $\mu$m region.  Both of these
features suggest the presence of a small amount of crystalline silicates (or
possibly different grain shapes or sizes) in the flared disk.  The ``knee'' near
11.3 $\mu$m was detected by \citet{przygodda03}.  They attribute the
changes in the shapes of the 10 $\mu$m silicate emission features
in T Tauri stars to grain growth.  Larger grains can enhance the emissivity
longward of the peak, but not shortward of the peak.  Thus GG Tau suggests
an additional broadening mechanism at work, perhaps an admixture of
crystalline silicates or quartz.

The 10 $\mu$m silicate feature in FN Tau is quite distinctive.  The contrast
of the 10 $\mu$m feature to the neighboring continuum is less than in the
three previous cases.  The feature has a flat top from 10 to 11.5 $\mu$m.
In addition, there is a small emission feature at 9.4 $\mu$m.  The overall
broadening of the silicate feature could be produced by differing shapes
and sizes of the dust grains.  However, the small peak at 9.4 $\mu$m
requires a differing chemical composition.  The feature appeared equally
prominent in both independent nod-position spectra of this object, the
spectral noise implied by the scatter of data points is small at these
wavelengths, and we have not seen it in any other objects, so we
believe this feature is real.

The sequence IP Tau -- FM Tau -- GG Tau -- FN Tau shows a trend of
decreasing contrast of the 10 $\mu$m silicate feature relative to the
5\,--\,8 $\mu$m continuum.  This sequence also shows an increasing
5\,--\,8 $\mu$m continuum relative to the stellar continuum, while
the contrast of the 10 $\mu$m silicate feature to the stellar continuum
is relatively constant (ranging from 7 to 10).  From this we infer that the
emission from the superheated flared disk is relatively constant, while
the emission from the optically thick accretion disk is increasing, causing
the feature contrast to diminish.

The 5\,--\,14 $\mu$m spectrum of CY Tau is quite different than the
above four stars.  The continuum is well elevated from the stellar
photosphere, indicating emission from the inner parts of the optically
thick accretion disk.  There is only a small bump near 10 $\mu$m.
We believe this arises from silicate grains. The ratio of flux attributable
to silicate grains to the stellar continuum at 10 $\mu$m is only about
1.7 here. This appears to be a case of the loss of small grains (or growth
to large sizes) in the superheated flared disk, while the optically thick
accretion disk is still quite prominent; the 8 $\mu$m flux is at least 3
times larger than the expected flux from the stellar photosphere.

Similar interesting fine-structure in the 10 $\mu$m silicate feature
has been seen in Herbig Ae/Be stars \citep{bouwman01, vanboekel03}
and in T Tauri stars \citep{przygodda03, alexander03, meeus03} and
interpreted variously as due to grain growth and the presence of crystalline
silicates and crystalline quartz (silica).
In a forthcoming paper we will use self-consistent disk models, exploring the
effects of grain shape, size, composition and crystallinity, to understand these spectra.

CoKu Tau/4 is unique in this group.  There is no evidence at all for
excess emission from the optically thick accretion disk in the 5\,--\,8 $\mu$m
range.  In fact, our measured flux is 10-20\% below the extrapolated stellar
continuum (Figs. 1 \& 2).  This is greater than our estimated photometric
uncertainty.  T Tauri stars are notoriously variable; this discrepancy could
be explained by variability since the 2MASS observations of this star date back to 1998.
The K-band measurement of CoKu Tau/4 reported by \citet{leinert93} was
30\% lower than the 2MASS measurement, so this degree of variability is quite
plausible.  Beyond 8 $\mu$m, there is increased emission, a broad peak 10\,--\,12
$\mu$m, a small dip at 12.5 $\mu$m, and a rapid rise at 13\,--\,14 $\mu$m which
continues up to about 20 $\mu$m, where the spectrum flattens to longer wavelengths.  
We attribute this excess emission above the stellar photosphere to emission from 
heated dust surrounding the star.  The structure in the spectrum described above is
indicative of optically thin emission from small silicate grains similar to
those found around FM Tau.  The shifting of peaks to longer wavelengths and
distortion of the spectral shape is primarily due to the much lower
temperature of the dust grains here.  In addition, there may be a
blackbody-like continuum contributing at the longer wavelengths.  The 10\,--\,20
$\mu$m color temperature of the dust excess here is 135 K, whereas it is
283 K in FM Tau.  For an optically thin cloud, the flux will be the
mass-weighted average over Planck functions of the various dust
temperatures times the silicate emissivity.  The silicate emissivity is 
less at 20 $\mu$m than it is at 10 $\mu$m, so the implied average 
dust temperatures are less than 135 K.  It is believed that the 20 $\mu$m 
silicate emissivity is less than 0.6 times the 10 $\mu$m emissivity 
\citep{draine84, simpson91}. This means the implied silicate 
temperature is below 123 K. The extremely low temperature found 
for CoKu Tau/4 indicates there is very little observable dust at temperatures 
higher than 123 K.  This implies there are very few small grains closer 
than about 10 AU from the star.

Dividing the CoKu Tau/4 excess by a black body of temperature 123 K
reveals the underlying silicate emissivity.  The 9.7 and 18 $\mu$m silicate
peaks are clearly present. The 9.7 $\mu$m peak shows no evidence for
the 11.3 $\mu$m knee; it is the narrowest feature in this group.  This
feature is narrower than the Trapezium, and matches quite well the
excess seen from dust surrounding the supergiant $\mu$ Cep
\citep{russell75}.  This narrow feature indicates the dust grains
must be $<$ 1 $\mu$m in radius, and they can have no
considerable icy coating.

CoKu Tau/4 has a SED of Class II and an age of 1 Myr \citep{kenyon95}, but is not a
classical T Tauri star since the H${\alpha}$ emission line equivalent
width is only 1.8-2.8 {\AA} \citep{cohen79, kenyon98}.  Thus it appears
to not be actively accreting, and the lack of an optically thick inner accretion
disk is not surprising.  Detectable dust grains are mostly excluded from the
central 10 AU of this system.  One possibility for this is the dust grains in
this inner region have grown to such a large size that they can no longer
be seen.  In a forthcoming paper we will develop
models which can explain this interesting spectrum.

\section{Conclusions}

The 5\,--\,30 $\mu$m spectra of classical T Tauri stars in the Taurus
dark clouds are dominated by three distinct components: the stellar photosphere,
usually relatively small, except in the case of CoKu Tau/4; an excess continuum
from 5 to 8 $\mu$m, associated with the inner parts of the optically thick
accretion disk; and emission from optically thin small silicate grains, dominating
the spectrum beyond 8 $\mu$m (except in the case of CY Tau).  These general
features are in accord with the theoretical models of these objects \citep{calvet92}.
The most prominent 10 $\mu$m features come from objects with the least
continuum excess (i.e. FM and IP Tau).  The relatively weak 10 $\mu$m
feature in CY Tau could indicate a lack of flaring in its disk (or grain growth
beyond 4 $\mu$m radius).

The detailed structure of the 10 $\mu$m silicate feature provides evidence
on grain processing in these objects.  The 9.7 and 18 $\mu$m emission
features in FM Tau indicate silicates which are most similar to those found 
throughout our galaxy in the interstellar medium, while the 11.3 $\mu$m 
``knee'' indicates some processing has occurred. IP and GG Tau exhibit more 
processing, as suggested by a more pronounced ``knee'' at 11.3 $\mu$m and 
a broader 10 $\mu$m feature, most likely explained by an admixture of 
modified grains.  FN Tau has the most complex 10 $\mu$m feature; 
in addition to the broadening noted above, there is a small peak at
$\sim$ 9.4 $\mu$m.

The spectrum of CoKu Tau/4 is perhaps the most interesting of this sample.  The
absence of an elevated 5\,--\,8 $\mu$m continuum is consistent with the lack of
accretion onto this star.  The very cold temperatures implied by the excess
emission indicates the absence of observable dust grains closer than  $\sim$
10 AU from the star.  We suggest CoKu Tau/4 is a transition object between a
classical T Tauri star and a weak-lined T Tauri star.  The remains of the
optically thick outer disk are still present, while the inner disk has been
extensively cleared.  Presumably, accretion through the outer disk is still
occurring, but something is happening to that material before it reaches the
star.  The 10 $\mu$m silicate feature is the narrowest of this sample.  This
indicates the dust in the outer parts of the disk has been little modified
from its origins in the ISM.

%\end{document}

\acknowledgments 
We thank our anonymous referee for a prompt and thorough review 
which improved our paper. This work is based on observations
made with the Spitzer Space Telescope, which is operated
by the Jet Propulsion Laboratory, California Institute of
Technology under NASA contract 1407. Support for this work
was provided by NASA through Contract Number 1257184 issued
by JPL/Caltech and through the Spitzer Fellowship Program,
under award 011 808-001.

\clearpage

\begin{deluxetable}{ccccccc}
\tablecaption{Facts about the T Tauri stars \label{table1}}
%\tablewidth{50pt}
\tablehead{
\colhead{Name} &\colhead{AOR key}  & \colhead{Sp} &\colhead{$A_{V}$} &\colhead{$T_{e}$} &\colhead{$\Omega_*$} &\colhead{$T_{color}$}\\
&\colhead{} &\colhead{} &\colhead{} &\colhead{} &\colhead{$10^{-19} str$} &\colhead{$10-20 \ \mu m$}}
\startdata
IP Tau & 3535616 & M0 & 0.2 & 3750 & 3.8 &\nodata \\
FM Tau & 3544832 & M0 & 0.7 & 3750 &2.7 & 283 \\
GG Tau & 3532032 & K7 & 0.8 & 3900 &8.5 &\nodata \\
FN Tau & 3534592 & M5 & 1.4 & 3200 &5.6 &\nodata \\
CY Tau & 3550208 & M1 & 0.1 & 3700 &3.0 &\nodata \\
CoKu Tau/4 & 3548416 & M1.5 & 0.9 & 3600 & 2.9 & 135 \\
\enddata

\tablecomments{The spectral type Sp, the visual extinction $A_{V}$ and the
effective temperature $T_e$ were taken from \citet{kenyon95}. $\Omega_*$ is the
solid angle a blackbody of temperature $T_e$ would have to match the 2MASS K-band
flux of the system.  The AOR key
 can be used to find details of the observations by
consulting the Observing Schedules available at the Spitzer website at
http://ssc.spitzer.caltech.edu/.}

\end{deluxetable}

\clearpage

\notetoeditor{Please display the two figures as one-column-wide
figures in the final (two-column) version of this paper.}

\begin{figure}[t]
%  \epsscale{0.75}
%  \plotone[angle=-90]{figure1.ps}
\includegraphics[angle=-90]{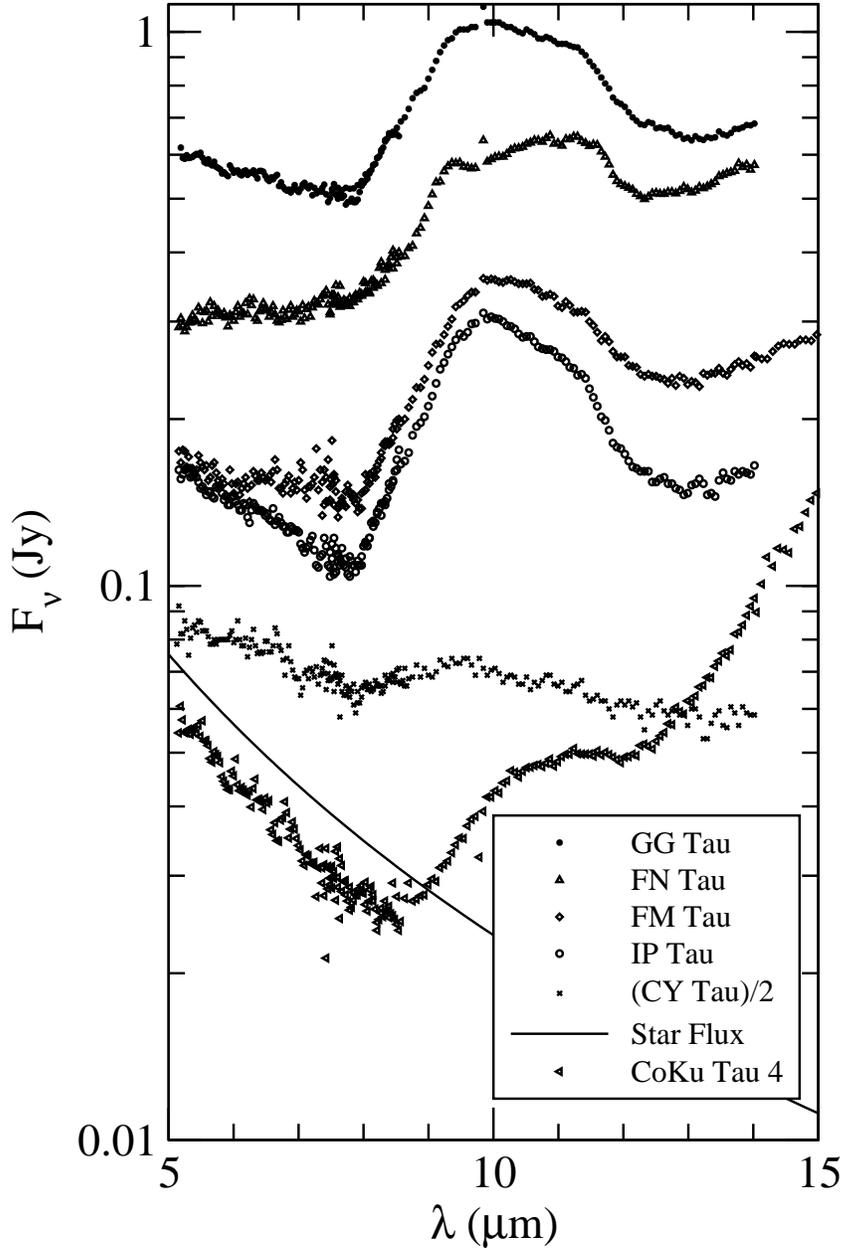}
  \caption{IRS spectra of GG Tau, FN Tau, FM Tau, IP Tau, CY Tau, and CoKu Tau/4 (in order, top to bottom).
           The spectrum of CY Tau has been divided by 2.  The smooth curve labelled
            ``Star Flux'' represents the stellar photosphere
             of both CoKu Tau/4 and FM Tau as described in the text.
          }
\end{figure}

\begin{figure}[t]
%  \epsscale{0.9}
%  \plotone[angle=-90]{figure2.ps}
\includegraphics[angle=-90]{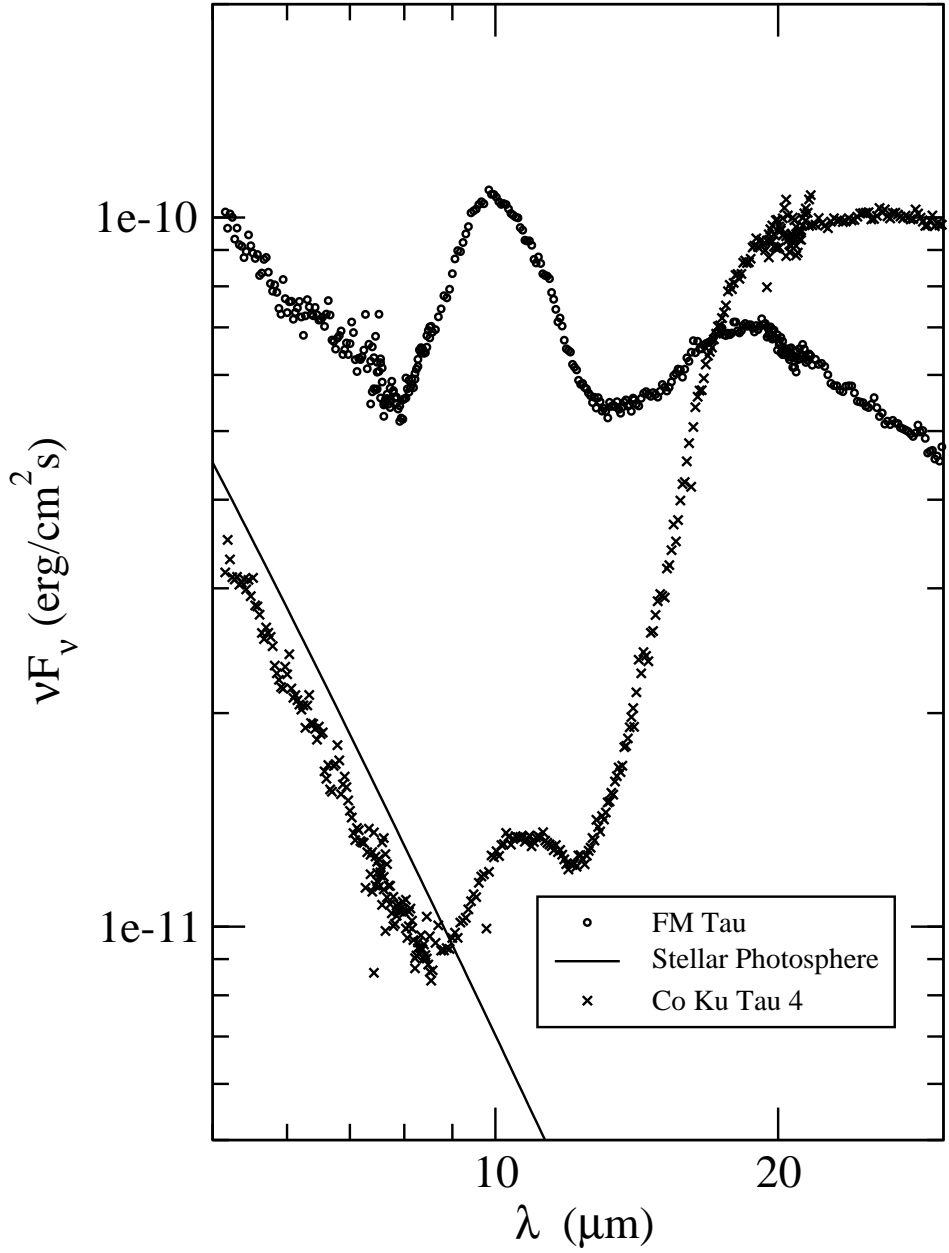}
 \caption{IRS spectra of FM Tau and CoKu Tau/4.
           The smooth curve labelled
            ``Stellar Photosphere'' represents the stellar photosphere
             of both CoKu Tau/4 and FM Tau as described in the text.
          }
\end{figure}

\end{document}